\documentclass[aps,prl,twocolumn,showpacs]{revtex4-1}
\usepackage{amsmath,amssymb}
\usepackage{graphicx}% Include figure files
\usepackage{color}
\begin{document}
\title{Parallel execution of quantum gates in a long linear ion chain \\via Rydberg mode shaping}
\author{Weibin Li$^1$, Alexander W. Glaetzle$^2$, Rejish Nath$^2$, and Igor Lesanovsky$^1$}
\address{$^1$School of Physics and Astronomy, The University of Nottingham, Nottingham, NG7 2RD, United Kingdom}
\address{$^2$Institute for Theoretical Physics, University of Innsbruck,\\
and Institute for Quantum Optics and Quantum Information of the Austrian Academy of Sciences, A-6020 Innsbruck, Austria}
\date{\today}
\begin{abstract}
We present a mechanism that permits the parallel execution of multiple quantum gate operations within a single long linear ion chain. Our approach is based on large coherent forces that occur when ions are electronically excited to long-lived Rydberg states. The presence of Rydberg ions drastically affects the vibrational mode structure of the ion crystal giving rise to modes that are spatially localized on isolated sub-crystals which can be individually and independently manipulated. We theoretically discuss this Rydberg mode shaping in an experimentally realistic setup and illustrate its power by analyzing the fidelity of two conditional phase flip gates executed in parallel. Our scheme highlights a possible route towards large-scale quantum computing via vibrational mode shaping which is controlled on the single ion level.
\end{abstract}
\maketitle

The ability to execute multiple quantum operations in parallel is believed to be a fundamental requirement for achieving large-scale quantum computation~\cite{divincenzo00,steane07,haffner08}. Among the many types of systems being considered for the physical implementation of a quantum processor~\cite{nielsen00}, trapped ions have attracted much attention for the astonishingly high degree of experimental control that can be gained over their internal and external degrees of freedom~\cite{haffner08}. One strategy for achieving parallelism is to build many local quantum processors. Current proposals envision setups where ions are confined in spatially separated wells provided by arrays of microtraps~\cite{cirac00} or by traps with segmented electric field electrodes~\cite{kielpinski02}. Few ions trapped within a given well form one of many local quantum processors that can be operated independently and in parallel~\cite{james98}. Information can be exchanged among different local processors by rearranging the potential landscape such that previously disconnected ions have common vibrational modes. Such rearrangement is usually achieved by switching voltages applied to the ion trap electrodes~\cite{kielpinski02}. In spite of the availability of microstructured arrays it remains a challenge to obtain fast switching times and a high spatial resolution of the local electric fields that would grant a manipulation of the potential landscape down to the level of a single ion. 

\begin{figure}[h!]
\includegraphics[width=3.38in]{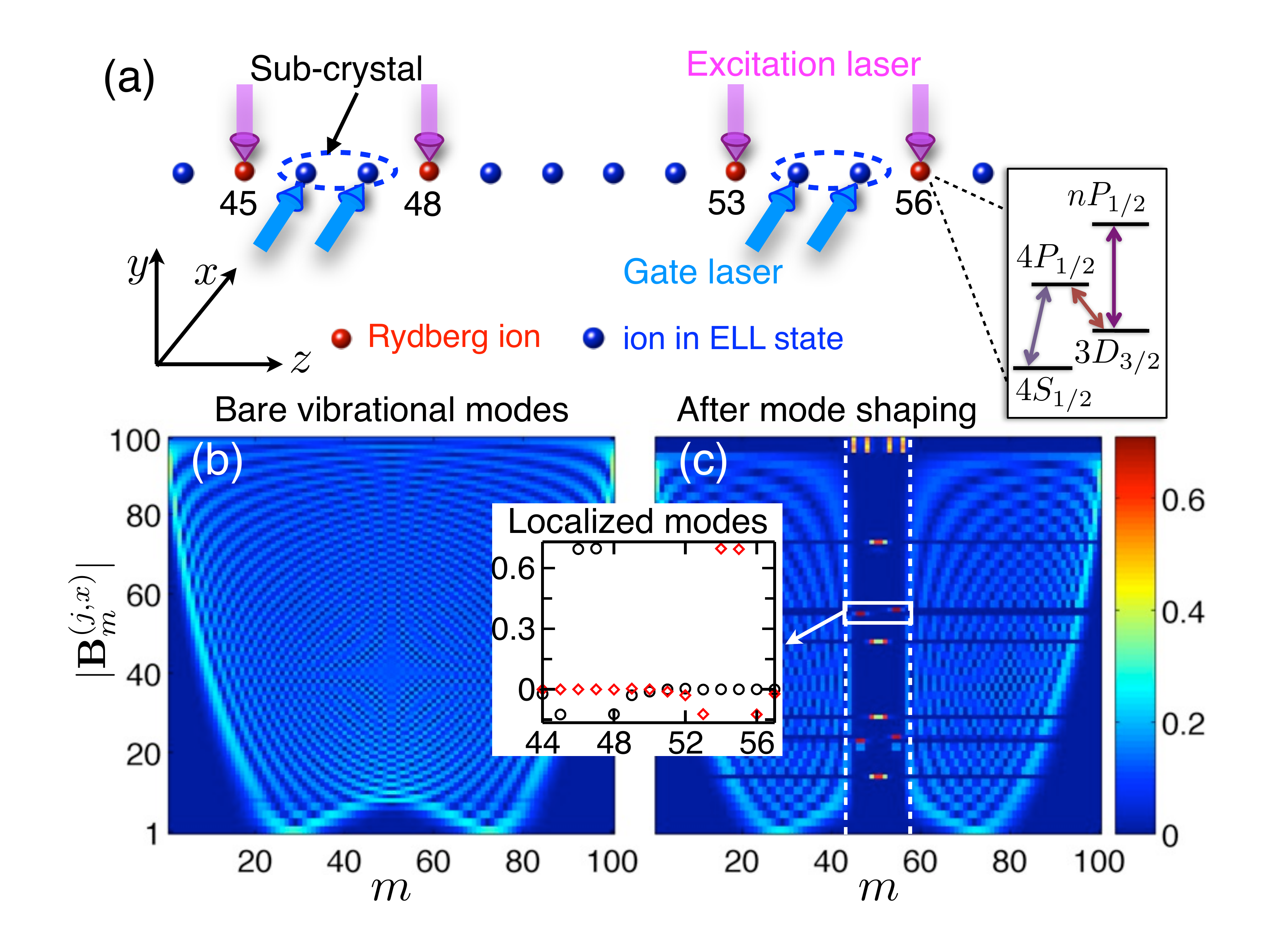}
\caption{(Color online) (a) Level structure of Ca$^+$ and schematics of the envisioned setup. Red and blue symbols refer to ions in Rydberg states and ions in electronically low lying (ELL) states, respectively. Sub-crystals of ion pairs are isolated within a linear crystal formed by 100 ions by the excitation of selected ions to the Rydberg $nP_{1/2}$-state (here the 45th/48th and 53rd/56th). Using laser-induced spin-dependent forces quantum gates can be executed on the two sub-crystals in parallel. (b) Vibrational modes of a crystal formed by 100 ions in ELL states. Depicted is the modulus of the normal mode matrix $\mathbf{B}_m^{(j,x)}$ where $j$ ($m$) refer to the mode (ion) index (see text for further detail). (c) Vibrational modes in the presence of four Rydberg ions. The white dashed lines delimit the region corresponding to the ions that are shown in panel (a). The Rydberg ions drastically reshape the vibrational mode structure leading to the emergence of modes that are localized on the two sub-crystals (see inset).}
\label{fig:fig1}
\end{figure}

In this work we introduce a scheme that permits the execution of multiple quantum gates in parallel on a long linear ion crystal. The method relies on the shaping of the vibrational crystal modes through the laser excitation of selected ions into electronically excited Rydberg states. Strong coherent forces acting on these excited ions~\cite{mueller08,li12} effectively break the long crystal into small sub-crystals in the sense that vibrational modes emerge which are strongly localized on only a few ions. We illustrate the power of this Rydberg mode shaping by thoroughly analyzing the fidelity of two two-qubit conditional phase flip (CPF) gates that are executed in parallel on different sub-crystals belonging to the same ion chain. A feature of our scheme is that decoupling between the localized modes and the remaining spectator modes permits us, not only to achieve a high gate fidelity, but also to drastically reduce the complexity of gate optimization protocols~\cite{garcia03,garcia05,duan04}. In view of the intrinsic stability of ion crystals~\cite{lin09}, the ability to address single ions individually by lasers~\cite{kaler11} and the long lifetime of Rydberg states, we believe that this approach indeed highlights a viable route towards achieving large-scale quantum computation.

Before providing details let us briefly outline the setup we have in mind. We consider a long linear crystal of $100$ $^{40}$Ca$^+$ ions which is realized within a quartic electric potential (see details below). This choice is motivated by the proposal discussed in Ref.~\cite{lin09} which envisages the implementation of an ion quantum processor where a long ion chain is divided into two parts: Quantum computation is carried out in the central region where ions are nearly uniformly spaced. The remaining outer ions are continuously Doppler cooled to prevent heating. To describe the internal structure of the ions we consider the four states depicted in Fig.~\ref{fig:fig1}a. The electronically low-lying (ELL) $S$, $P$ and $D$ states are employed in numerous ion trap experiments for the storage, manipulation and read out of quantum information~\cite{haffner08}. Furthermore, we consider the Rydberg state $nP_J$ (with $J=1/2$ and the principal quantum number $n$) which is excited from the $3D_{3/2}$-state via a single photon transition~\cite{kaler11,kolbe12}.

We envisage Rydberg excitations to be carried out in the central region of the ion chain and in Fig.~\ref{fig:fig1}a we illustrate a situation where four Rydberg ions enclose two pairs of ions in ELL states. Those ion pairs will form the sub-crystals on which we are going to execute quantum gates in parallel. The underlying physical mechanism which we aim to exploit for this purpose becomes apparent in Figs.~\ref{fig:fig1}b,c. Here we show the absolute values of the normal mode matrix of the vibrational crystal modes which provide a measure on how much each ion contributes to a vibrational mode. In Fig.~\ref{fig:fig1}b, that shows the case in which all ions are in ELL states, we see that in general many ions contribute to each normal mode. Compared to this, the presence of Rydberg ions leads to a drastic change of the mode structure as can be seen in Fig.~\ref{fig:fig1}c. The reason is rooted in the large polarizability $\mathcal{P}_{nP}$ of Rydberg states~\cite{li12} which modifies the local trapping potential, leading essentially to a constriction of the ion chain at positions where Rydberg ions are excited. The selective Rydberg ion excitation thus creates localized modes, primarily occupying the two isolated sub-crystals composed by ions in ELL states. This Rydberg mode shaping permits the parallel execution of quantum gates on the two sub-crystals. This is similar in spirit to the idea underlying segmented ion traps~\cite{harlander11,brown11}. The advantage of our approach is that due to the availability of single ion laser addressing it fundamentally permits the control of the potential landscape on the smallest achievable length scale - namely on the level of single ions.
\\
\noindent{\it Long linear ion chain - }
Let us now provide a more detailed discussion of the practical implementation of the above idea. To achieve a long ion crystal we consider an ion trap formed by the time-dependent electric potential $\Phi(\mathbf{r},t)=\Phi_\mathrm{rf}(\mathbf{r},t)+\Phi_\mathrm{st}(\mathbf{r})$.
Here $\Phi_\mathrm{rf}(\mathbf{r},t)=\alpha\cos\Omega t(x^2-y^2)$ is the potential of a radio-frequency (rf) field with gradient $\alpha$ and frequency $\Omega$ and $
\Phi_\mathrm{st}(\mathbf{r})=\beta_2(2z^2-r^2)/2+\beta_4 \left[z^4-3z^2 \,r^2+3r^4/8\right],
$ with $r^2=x^2+y^2$, is a quartic static electric potential whose parameters $\beta_j~(j=2,4)$ depend on the specifics of the field generating electrodes, i.e. the gradient and higher derivatives of the field. Recently similar potentials have been realized experimentally~\cite{harlander11,brown11}. For a sufficiently fast rf frequency drive~\cite{leibfried03} an ion of mass $M$ experiences the ponderomotive potential
$V_{\mathrm{p}}(\mathbf{r})=e\left[\frac{e\alpha^2}{M\Omega^2}r^2+\Phi_\mathrm{st}(\mathbf{r})\right]$. Within this trap an ion chain is formed along the $z$-axis provided that
$\alpha\gg \{|\beta_2|,\beta_4l_s^2\}>0$ ($\beta_2<0$). Here $l_s=[e/(8\pi \epsilon_0 |\beta_2|)]^{1/3}$ is the typical length scale associated with $V_{\mathrm{p}}(\mathbf{r})$ and $e$ and $\epsilon_0$ are the elementary charge and the vacuum permittivity, respectively. In addition, the tuning of the parameters $\beta_2$ and $\beta_4$ permits us to achieve a long ion crystal in which the equilibrium positions of ions in ELL states are approximately evenly spaced~\cite{lin09}. The equilibrium positions of the long ion crystal is determined by a parameter $k_4=2\beta_4l_s^2/|\beta_2|$. In the following we set $k_4=1.343$ as this choice minimizes fluctuations of the nearest neighbor separation within the central region of the ion chain~\cite{lin09}.

In Ref.~\cite{li12} we showed that ions excited to the $nP_{1/2}$-Rydberg state experience not only the ponderomotive potential but are also subject to an additional radial potential that is proportional to the Rydberg polarizability $V_{\mathrm{a}}(\mathbf{r})\approx -e^2\alpha^2 \mathcal{P}_{nP} r^2$, where $\mathcal{P}_{nP}\approx -0.25\times n^7$ (in atomic units). There are also small corrections to the trapping potential along the $z$-axis but those are negligible in this linear ion trap. The ratio of the radial trap frequencies experienced by an ion in the Rydberg/ELL state is approximately given by $\omega_\mathrm{Ryd}/\omega_\mathrm{ELL}=\sqrt{1-M\Omega^2\mathcal{P}_{nP}}$. In practice ratios on the order of $2$ and larger can be achieved. In the following we will show that this is already sufficient for a Rydberg ion to effectively introduce a constriction of the linear ion chain which strongly affects the vibrational mode structure.

\noindent{\it Collective modes and mode shaping -} Let us now demonstrate the mode shaping considering the transverse phonon modes along the $x$-axis as an example. The treatment of the $y$-phonons is done accordingly. The phonon Hamiltonian is~\cite{james98}
$H_{\mathrm{v}}=\sum_{j=1}^{N}\hbar\omega_j(b^{\dagger}_jb_j+1/2)$. Here $b^{\dagger}_j$ ($b_j$) is the creation (annihilation) operator of the $j$-th phonon, whose frequency $\omega_j$ is calculated by diagonalizing the Hessian matrix ($\sum_m\mathcal{H}_{mn}\mathbf{B}_m^{(j,x)}=(\omega_j/\omega_s)^2\mathbf{B}_n^{(j,x)}$) with
\begin{equation}
\mathcal{H}_{mn}=\left\{\begin{array}{ll}\left[\frac{\omega_m^{(x)}}{\omega_s}\right]^2+\frac{1}{2}-\frac{3k_4}{2}z_m^2-\sum\limits_{k\neq m}^N\frac{1}{|z_k-z_m|^3}, & n=m \\ \frac{1}{| z_m-z_n|^3},&n\neq m \end{array}\right.
\nonumber
\end{equation}
and $\mathbf{B}_m^{(j,x)}$ denoting the eigenvectors. The parameters $z_m$ are the $z$-component of the equilibrium position of the $m$-th ion and   $\omega_m^{(x)}$ is the state-dependent trapping frequency of the $m$-th ion along the $x$-axis, i.e. $ \omega_m^{(x)}=\omega_{\mathrm{Ryd}} (\omega_{\mathrm{ELL}})$ if the ion is in the Rydberg (ELL) state. For convenience, we have defined a reference frequency $\omega_s=\sqrt{2e|\beta_2|/M}$ and scaled length with $l_s$ such that the Hessian is dimensionless.

\begin{figure}[h]
\includegraphics[width=3.3in]{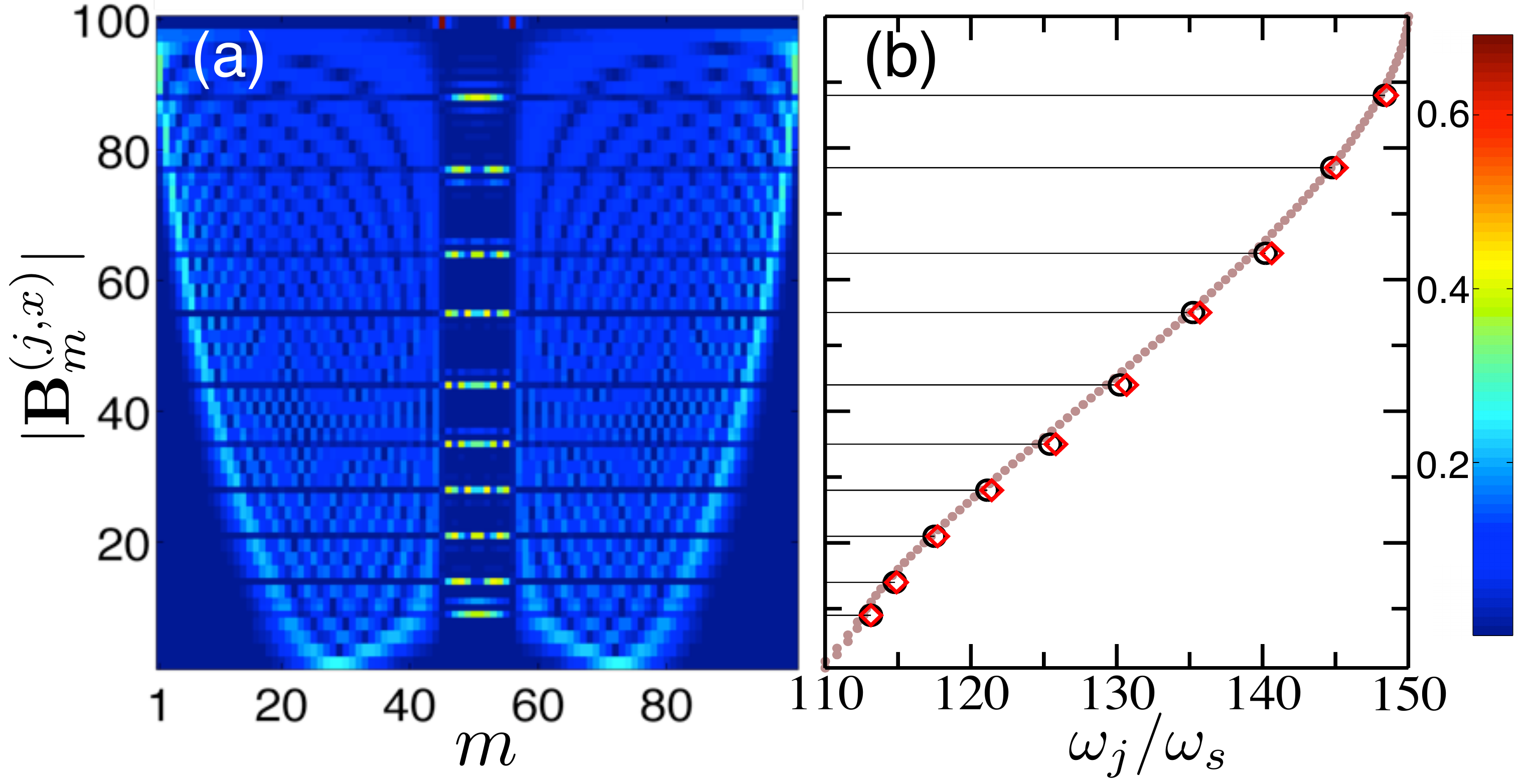}
\caption{(Color online) (a) Modulus of the eigenvector $\mathbf{B}^{(j,x)}_m$ when ions 45 and 56 are in the Rydberg state. Localized modes reside on ions in ELL states forming the sub-crystal delimited by the two Rydberg ions. Panel (b) shows the corresponding eigenenergy of the localized modes obtained from the full (circle) and truncated (diamond) calculations. The largest discrepancy between these calculations is about $0.3\%$. The dots show the (quasi-continuous) energy spectrum of the ion chain without mode shaping. In all the calculations we use $\omega_{\mathrm{ELL}}/\omega_s=150$ and $\omega_{\mathrm{Ryd}}/\omega_s=198.5$. }
\label{fig:fig2}
\end{figure}

We start with a simple situation where in our chain of 100 ions the 45th and 56th are excited to the Rydberg state. The resulting change of the mode structure becomes directly apparent in the modulus of $\mathbf{B}^{(j,x)}_m$ which is depicted in Fig.~\ref{fig:fig2}a. Compared to the situation without mode shaping (Fig.~\ref{fig:fig1}b), the striking difference is that the 46th to 55th ion constitute a virtually isolated sub-crystal hosting a series of \emph{spatially localized modes}. The energies of these local modes are shown in Fig.~\ref{fig:fig2}b where we also undertake a comparison to the mode energies obtained by considering exclusively the sub-crystal, i.e., a truncated linear crystal composed out of only 10 ions. Note, that although we are considering here a case in which the sub-crystal ions are symmetrically positioned around the center of the ion crystal, our observations remain true also in asymmetric situations. To effectively create sub-crystals with localized modes we require that $(\omega_{\mathrm{Ryd}}/\omega_s)^2\gg \max(\mathcal{H}_{mn})$ $(m\neq n)$ with $\max(\mathcal{H}_{mn})$ being the maximum of the off-diagonal matrix elements of the Hessian. This condition means that the energy of vibrational modes to which Rydberg ions participate significantly is much larger than the energy of the collective modes of ions in ELL states.

\noindent{\it Parallel CPF gates -}
Let us return to this initial example in which we had two sub-crystals composed by the ion pairs $\{46,47\}$ and $\{54,55\}$. Each of the sub-crystals hosts two localized vibrational modes. The eigenvector corresponding to the localized mode with higher energy is displayed in the inset of Fig.~\ref{fig:fig1}c. In the following we will show that with these local modes, we can execute two two-qubit gates in parallel. Specifically, we discuss a $\sigma^z$-type~\cite{schneider12} two-qubit CPF gate. Qubits are encoded in two ELL states of an ion, denoted by $|\uparrow\rangle$ and $|\downarrow\rangle$. These can be hyperfine states as discussed in Refs.~\cite{leibfried03gate,lee05}, or states coupled by optical quadrupole transitions as, e.g., in Ref.~\cite{kim08}. The CPF gate is implemented by a laser induced coupling (see gate lasers in Fig. \ref{fig:fig1}a) between the qubit states and the vibrational crystal modes/phonons. This results in a `spin-dependent' force~\cite{leibfried03gate,lee05,kim08} whose action is described by the spin-phonon Hamiltonian~\cite{zhu06,garcia05} $
H_\mathrm{I}=\sum_{m,j=1}^{N}\hbar\Omega_m(t)\sigma_m^z\eta_m^{(j)}\mathbf{B}_m^{(j,x)}(b_j^{\dagger}e^{i\omega_j t}+\mathrm{h.c.})
$. Here $\Omega_m(t)$ is the time-dependent Rabi-frequency of the gate laser that addresses the $m$-th ion and $\eta_m^{(j)}=k\,l_j$ is the corresponding Lamb-Dicke parameter, with $k$ being the modulus of the laser wave vector and $l_j=\sqrt{\hbar/2M\omega_j}$ being the oscillator length associated with the $j$-th phonon mode. The CPF gate is conducted by switching the gate lasers on for a given time $\tau$ during which the Rabi frequencies $\Omega_m(t)$ are varied. Using the Magnus formula~\cite{blanes09}, the evolution operator due to $H_\mathrm{I}$ is then given by
\begin{equation}
\label{eq:rydbergphonon_evolution}
U(\tau)=\exp\left[i\sum_{m}Q_m(\tau)\sigma_m^z+i\sum_{mn}\phi_{mn}(\tau)\sigma_m^z\sigma_n^z\right].
\end{equation}
The first term in the exponential characterizes the residual coupling of the $m$-th qubit with the phonon modes and depends on $Q_m(\tau)=\sum_j[\alpha_m^{(j)}(\tau)b_j^{\dagger}+\mathrm{h.c.}]$ where $\alpha^{(j)}_m(\tau)$ is a parameter that characterizes the coupling strength. The second term gives rise to a phonon-induced spin-spin coupling between the $m$-th and $n$-th qubit thereby effectuating a CPF gate. A perfect CPF gate is realized when $\phi_{mn}(\tau)=\pi/8$ and $\alpha_m^{(j)}(\tau)=0$. As shown in Refs. \cite{garcia03,garcia05,duan04} this can be achieved via optimizing the time-dependent profile of the Rabi frequencies $\Omega_m(t)$. Such optimization is challenging since in general many phonon modes contribute even when only a single CPF gate operation is conducted within a long ion chain. Rydberg mode shaping has the potential to drastically reduce the complexity of such optimization procedure as even in long crystals only few vibrational modes actually couple to the qubit ions located on a sub-crystal.

Let us now analyze the performance of two CPF gates that are executed in parallel on the two sub-crystals depicted in Fig.~\ref{fig:fig1}a. To assess the performance of the gate operation, we use the high energy mode (inset of Fig.~\ref{fig:fig1}c) as `quantum bus'. The gate lasers are switched on for a time $\tau=8\tau_{\mathrm{b}}$, where $\tau_{\mathrm{b}}=2\pi/\omega_{\mathrm{b}}$ is the oscillation period of the bus mode and the Rabi frequency is assumed to follow $\Omega_m(t)=\Omega_0\sin(\nu t)$ (as also discussed in Ref.~\cite{lee05}). We optimize the fidelity with respect to the parameter $\nu$ of this simple ansatz. Imposing a two-qubit phase shift $\phi_{mn}(\tau)=\pi/8$ fixes the value of the amplitude $\Omega_0$ (for more detail see Ref.~\cite{zhu06}). The qubits are prepared in a product state $|\Psi(0)\rangle= \left(|\psi_{m_1}\rangle\otimes|\psi_{n_1}\rangle\right)\otimes\left(|\psi_{m_2}\rangle\otimes|\psi_{n_2}\rangle\right)$ with $|\psi_m\rangle=(|\uparrow_m\rangle+|\downarrow_m\rangle)/\sqrt{2}$ and $\{m_j,n_j\}$ being indices of ions forming the $j$-th sub-crystal. Ideally, the output state after the parallel execution of the two CPF gates is
$|\Psi(\tau)\rangle=\exp[i\pi/4(\sigma_{m_1}^z\sigma_{n_1}^z+\sigma_{m_2}^z\sigma_{n_2}^z)]|\Psi(0)\rangle$.
However, due to the residual phonon-qubit coupling this state will be only reached with a certain probability, which we characterize through the fidelity $F=\langle\Psi(\tau)|\mathrm{Tr}_{\mathrm{v}}\rho(\tau)|\Psi(\tau)\rangle$. Here, $\rho(t)=U(t)\rho(0)U^{\dagger}(t)$ with $\rho(0)=\rho_{\mathrm{v}}\otimes|\Psi(0)\rangle\langle\Psi(0)|$ and $\mathrm{Tr}_{\mathrm{v}}$ denotes the trace over the vibrational modes whose density matrix is $\rho_{\rm{v}}$. For calculating the fidelity we assume the following sequence: The ions are initially in ELL states and the phonon density matrix $\rho_{\rm{v}}$ is a thermal distribution. Rydberg ions are subsequently excited via protocol that is highly non-adiabatic w.r.t. the phonons, i.e. the phonon density matrix is unchanged. In the Supplementary material - Part A we provide more detail on such protocol. The gate fidelity is then calculated via a transformation that expresses $\rho_{\rm{v}}$ in terms of the shaped vibrational modes~\cite{li12} (see Supplementary material - Part B).

\begin{figure}[h]
\includegraphics[width=3.43in]{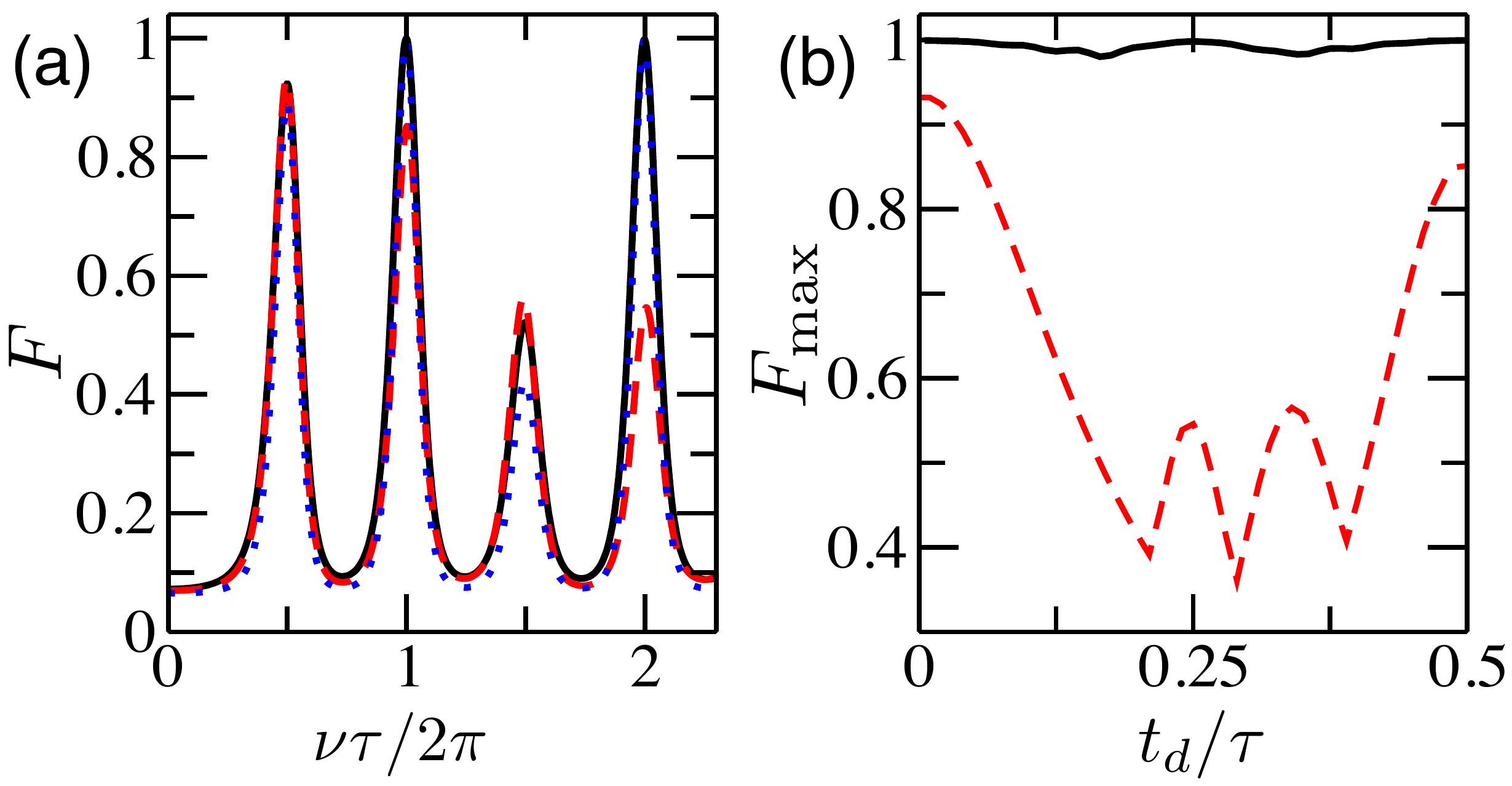}
\caption{(Color online) (a) Fidelity of the two CPF gates. The solid (dotted) curve is the result calculated using all (only the four localized) modes. The dashed curve corresponds to the gate fidelity without mode shaping, where the bus mode is the highest energy mode (Fig.~\ref{fig:fig2}b), whose average phonon number is $3.25$. In the calculations, we assume that all phonon modes when the ions are in the ELL states have the same temperature. (b) Maximal fidelity vs. delay time. $F_{\mathrm{max}}$ is found by maximizing the fidelity over $\nu$ within the range shown in (a). The solid (dashed) curve stands for the calculation with (without) mode shaping.}
\label{fig:fig3}
\end{figure}

Let us first consider a situation in which the two CPF gates start simultaneously. We find that the highest achievable fidelity within our simple ansatz is $F_{\mathrm{max}}\approx99.95\%$. As shown in Fig.~\ref{fig:fig3}a, this maximum occurs at $\nu\tau=2\pi\times K$ (with $K$ an integer). At these points the bus modes almost entirely return to their initial states~\cite{garcia05,sorensen99}. The fact that such a high fidelity is achievable within this simple ansatz is a direct consequence of the fact that the Rydberg ions delimiting the sub-crystal lead to a dramatic reduction of the number of vibrational modes that couple to the qubit ions. Without this Rydberg mode shaping the highest fidelity that we can achieve is $93\%$.

The power of the mode shaping becomes even more apparent when introducing a start time delay $t_\mathrm{d}$ of the second CPF gate with respect to the first one. $F_{\mathrm{max}}$ slightly decreases with growing $t_\mathrm{d}$ but always remains above $98\%$ as shown in Fig.~\ref{fig:fig3}b. This demonstrates that the two CPF gates can be operated essentially independently. In the absence of mode shaping, however, $F_{\mathrm{max}}$ quickly drops with increasing $t_\mathrm{d}$ reaching a minimal value of $\approx36.1\%$.

Finally, let us discuss additional sources that would influence the gate fidelity. Firstly, the fidelity will be in principle reduced by the radiative decay of the Rydberg state. This is mitigated by the fact that CPF gates are inherently fast~\cite{zhu06} and that Rydberg states are long lived. For example, choosing the $60P_{1/2}$ state, we obtain a Rydberg lifetime $\approx 270~\mu$s and with the trap parameters $\alpha=7\times10^8\ \text{V/m}^2$, $\Omega= 2\pi\times 25.2$ MHz and $\beta_2=-2.09\times10^3\ \text{V/m}^2$ we obtain a gate time $\tau\approx 3.7~\mu$s. Taking into account the Rydberg excitation time, the gate fidelity will be modified by an overall factor, about $0.982^{N_{\mathrm{Ryd}}}$ with $N_{\mathrm{Ryd}}$ the number of Rydberg ions that are excited during the gate operation. However, in principle the gate operations can be accelerated by using a more sophisticated optimization protocol~\cite{lin09}, and furthermore optimized gate schemes can be imagined in which the Rydberg ions do not stay permanently excited. Secondly, infidelities are caused by other factors, such as the anharmonicity of  the ionic motion and corrections beyond the Lamb-Dicke limit. These have been investigated in detail by Lin {\it et al.} in Ref.~\cite{lin09} and their contributions have been found to be marginal.

In conclusion, we showed that the transverse vibrational modes of a linear ion chain can be shaped by the selective excitation of Rydberg ions leading to the emergence of strongly localized modes. Rydberg mode shaping, when applied to larger (higher dimensional) ion crystals, might indicate a route towards scalable quantum computation within a single large ion crystal. One can think of dedicated ions which are not used as qubits but only for segmenting the large crystal. When excited they give rise to local modes that permit the parallel manipulation of sets of qubits. When deexcited the non-local character of the vibrational modes is restored permitting the entanglement of more distant qubits.

\acknowledgements
We thank Peter Zoller for initial discussions which stimulated this work. We thank C. Ates and S. Genway for careful reading of the manuscript. Discussions with all members of the R-ION consortium are kindly acknowledged. This work is funded through EPSRC and the ERA-NET CHIST-ERA (R-ION consortium).
\setcounter{figure}{0}
\setcounter{table}{0}
\setcounter{equation}{0}
\renewcommand{\thefigure}{S\arabic{figure}}
\renewcommand{\thetable}{S\Roman{table}}
\renewcommand{\theequation}{S\arabic{equation}}

\section{Supplementary Material for ``Parallel execution of quantum gates in a long linear ion chain via Rydberg mode shaping"}
\subsection{Part A - Excitation of Rydberg ions}
For the practical application of the mode shaping it is crucial to excite ions to the Rydberg state with high fidelity which can be a challenge if the trapping potentials of ELL and Rydberg states are very different: In general the (reduced) density matrix of the ion to be excited is given by $|3D_{3/2}\rangle\! \langle 3D_{3/2}|\otimes \rho_{D}$, where $\rho_{D}$ is the density matrix of the external degrees of freedom. The Rydberg excitation needs to take the ion to the state $|nP_{1/2}\rangle\! \langle nP_{1/2}|\otimes \rho_{P}$ as the shaped modes are only present if the electronic population of the ion is entirely transferred to $|nP_{1/2}\rangle$. Note, that the density matrix of the external degrees of freedom can in general change during the excitation process, e.g., due to non-trivial Franck-Condon factors which arise from the different potential shapes in low-lying and Rydberg states~\cite{li12}. There are a number of strategies to perform the transfer with high fidelity: For example, one can cool all phonon modes to the ground state, such that the initial state $\rho_{D}$ is precisely known. The excitation laser pulse (frequency, strength and duration) can then be optimized to achieve perfect electronic state transfer. Alternatively, one can use a broad band laser excitation that does not resolve the individual phonon modes \cite{poyatos96} and thus performs the transfer independently of the phonon state. Both methods, however, are rather challenging with current technology ~\cite{leibfried03,kaler11,kolbe12}. A currently feasible alternative is to remove the difference in the trapping potentials during the excitation process by altering the polarizability of Rydberg states through the application of a microwave field (MW) which creates dressed states of tuneable polaizability. The Rydberg excitation is then no longer different than the excitation of ELL states. After the excitation has been carried out the MW is switched off in a way that is adiabatic on electronic timescales but can be highly non adiabatic on the timescale of the phonon dynamics.

The scheme works as follows: Together with Rydberg laser, we apply a MW field that couples the Rydberg $|nP_{3/2}(1/2)\rangle$  (denoted  by $|P\rangle$) with a nearby s-state $|n'S_{1/2}(1/2)\rangle$ (denoted by $|S\rangle$). The corresponding ion-field interaction is given by
\begin{equation}
V(t)=-eE_0\cos\omega_0 t\, z-eE_1\cos\omega_1t\, z,
\end{equation}
where $E_0$ $(E_1)$ is the laser (MW) electric field and  $\omega_0$ ($\omega_1$) is the Rydberg laser (MW) frequency. Both the Rydberg laser and the MW field are linearly polarized along the $z$-axis. The field-free Rydberg energies are $\epsilon_S$ and $\epsilon_P$, respectively. To be concrete we also assume $\epsilon_S<\epsilon_P$. To proceed it is convenient to transform into the interaction picture. Using the unitary operator $U_{\rm{i}}=|D\rangle \langle D|+e^{i\omega_0 t}|P\rangle \langle P|+e^{i(\omega_0-\omega_1) t}|S\rangle \langle S| $ and in rotating-wave approximation, we obtain the Hamiltonian ($\hbar= 1$)
\begin{eqnarray}
\label{eq:ion_field}
H&=&\Delta_S|S \rangle \langle S|+\Delta_P|P\rangle \langle P|+H_{\rm{L}}, \\
H_{\rm{L}}&=& \frac{\Omega_{\rm{L}}}{2}(|P\rangle\langle D|+h.c) + \frac{\Omega_{\rm{MW}}}{2}(|S\rangle\langle P|+h.c.),\nonumber
\end{eqnarray}
where $\Delta_S=\epsilon_S-(\omega_0-\omega_1)$ and $\Delta_P=\epsilon_P-\omega_0$. $\Omega_{\rm{L}}=-eE_0\langle P|z|D\rangle$ and $\Omega_{\rm{MW}}=-eE_1\langle S|z|P\rangle$.
\begin{figure}[h]
\centering
\includegraphics[width=2.6in]{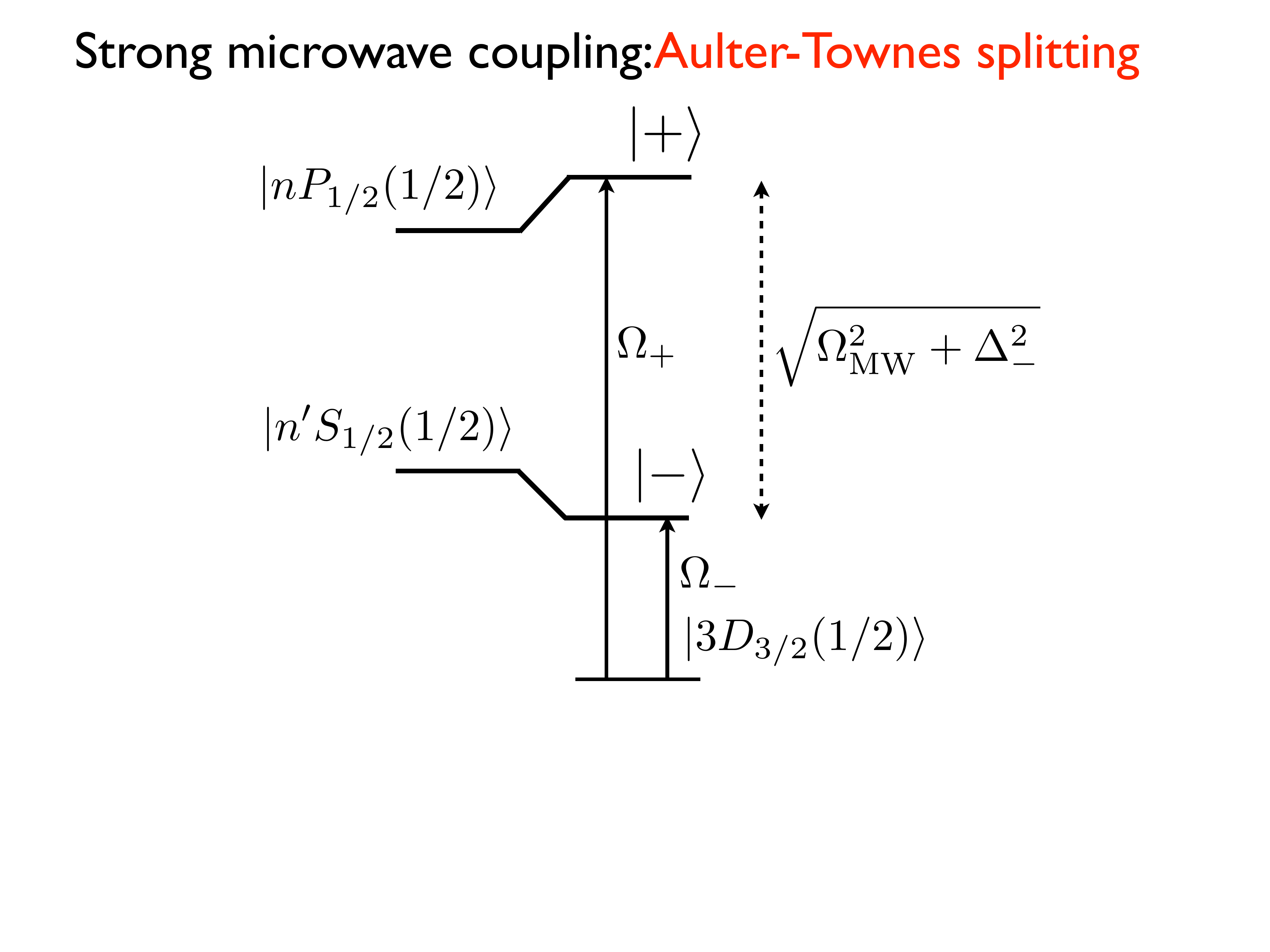}
\caption{Level scheme used in the Rydberg excitation. A strong MW field results in a large Autler-Townes splitting~\cite{autler55} between the two dressed states. }
\label{fig:suppfig1}
\end{figure}

We consider a strong MW field, $\Omega_{\rm{MW}}\gg \Omega_{0}$, for which it is convenient to use the dressed state in order to describe dynamics of the Rydberg states. By diagonalizing the part of Hamiltonian (\ref{eq:ion_field}) that contains the MW coupling part, the dressed states are given by
\begin{eqnarray}
|\pm\rangle&=&N_{\pm}\left(C_{\pm}|P\rangle+|S\rangle\right),
\end{eqnarray}
where $C_{\pm}=\frac{\Delta_-\pm\sqrt{\Omega_{\rm{MW}}^2+\Delta_-^2}}{\Omega_{\rm{MW}}}$ with $\Delta_{\pm}=\Delta_P\pm\Delta_S$ and $N_{\pm}$ is the normalization constant. The dressed state energy is $E_{\pm}=\frac{\Delta_+}{2}\pm\frac{1}{2}\sqrt{\Omega_{\rm{MW}}^2+\Delta_-^2}$. With the dressed state at hand, the Hamiltonian Eq.~(\ref{eq:ion_field}) becomes
\begin{eqnarray}
\label{eq:dressed_ion}
H &\approx& E_+|+\rangle\langle+|+E_-|-\rangle\langle-|+H_{\rm{L}}',\\
H_{\rm{L}}'&=&-\frac{1}{2}(\Omega_-|-\rangle\langle D|+h.c.)+\frac{1}{2}(\Omega_+|+\rangle\langle D|+h.c.),\nonumber\\
\end{eqnarray}
with $\Omega_{\pm}=\frac{\Omega_{\rm{MW}}}{2\sqrt{\Omega_{\rm{MW}}^2 +\Delta_-^2}N_{\pm}}\Omega_{\rm{L}}$.
Thus the low-lying $D$ state is now coupled with the two dressed state. Here the large energy splitting allows us to address the dressed states individually with the Rydberg laser.

\begin{figure}[h]
\centering
\includegraphics[width=3.2in]{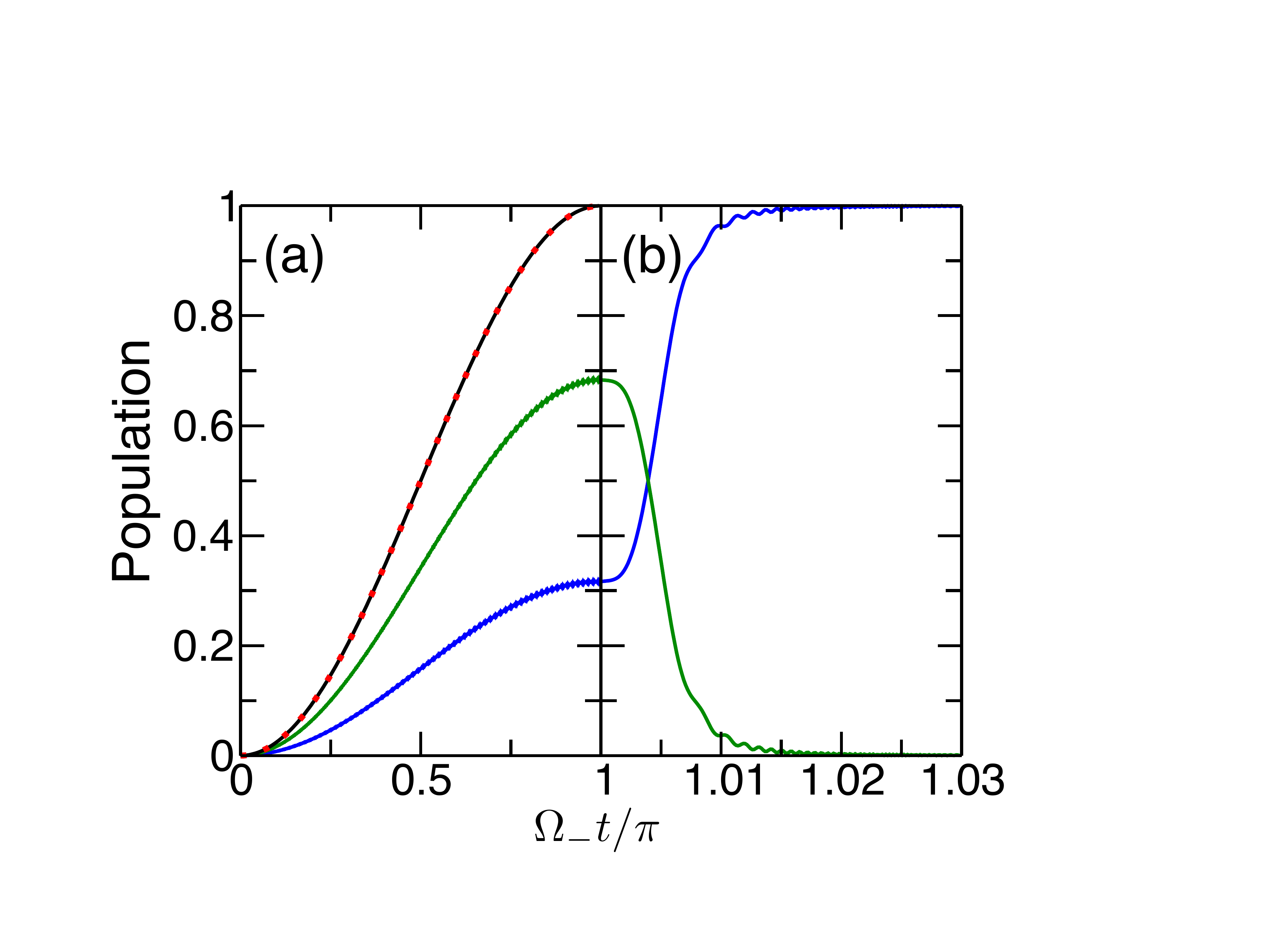}
\caption{(a) Carrier resonant transition of the $|-\rangle$ state. The solid (dotted) curves are solutions of the Hamiltonian Eq.~(\ref{eq:ion_field}) [Hamiltonian Eq.~(\ref{eq:dressed_ion})]. The blue and green curve correspond to the probability of the $|P\rangle$ and $|S\rangle$ state. The parameters are: $\Omega_-=2\pi\times 1$ MHz, $\Omega_{\rm{MW}}=2\pi\times 400$ MHz, $\Delta_S=2\pi\times 136.074$ MHz and $\Delta_P=2\pi\times 293.957$ MHz. These parameters lead to a vanishing polarizability of the dressed $|-\rangle$ state. (b) Adiabatic population evolution of the $P$ and $S$ state. The change rate is $c=\Omega_{\rm{MW}}/4.7$. }
\label{fig:suppfig2}
\end{figure}

The polarizability of the dressed state is $\mathcal{P}_{\pm}= N_{\pm}^2(C_{\pm}^2\mathcal{P}_{nP}+\mathcal{P}_{n'S}$). As $\mathcal{P}_{n'S}>0$ for high-lying Rydberg state, the polarizability of the dressed state vanishes under certain conditions. For example, for $n'=n$, $\mathcal{P}_{\pm}=0$ when $|C_{\pm}|\approx 0.68$, which can be realized by controlling the MW frequency and/or Rabi frequency.  On the other hand, when $\mathcal{P}_{\pm}=0$, the trapping potential of the Rydberg ion in the dressed state becomes identical with that of the ions in ELL states. In this case the Franck-Condon factors \cite{li12} become trivial and the laser excitation is no different to transitions driven among ELL states. In  Fig.~\ref{fig:suppfig2}a we demonstrate the $|-\rangle$ state excitation. After a $\pi$-pulse, the ion is excited to the $|-\rangle$ state.

Once the state $|-\rangle$ is excited, the Rydberg laser is switched off and also the MW is switched off such that $|-\rangle$ is adiabatically transferred to the state $|P\rangle$. As an example, we show in Fig.~\ref{fig:suppfig2}b a case in which the MW detuning is changed according to $\Delta_{SP}(t)=(\Delta_S-\Delta_P)[1-c^2t^2]$. The Rydberg ion is fully populating the $P$ state after about 13 nanoseconds.

\subsection{Part B - Calculating the gate fidelity}
As shown in the previous section the Rydberg excitation can be performed on a timescale that is highly non-adiabatic with respect to the vibrational dynamics. In the extreme case the phonon state does not change, i.e. is frozen, in the course of the Rydberg excitation. In this section, we show how for this case the gate fidelity can be calculated by using a Duschinsky transformation that connects the phonon modes before the Rydberg excitation (bare phonon modes) to the shaped phonon modes.

Let us first describe the normal coordinate $\mathcal{Q}_g$ and canonical momentum $\mathcal{P}_g$ of the bare mode using the corresponding phonon creation and annihilation operators~\cite{james98},
\begin{eqnarray}
\label{seq:bareQ}
\mathcal{Q}_g&=&\mathbf{L}_g(\mathcal{A}_{\dagger}+\mathcal{A}),\\
\label{seq:bareP}
\mathcal{P}_g&=&\mathbf{P}_g(\mathcal{A}_{\dagger}-\mathcal{A}),
\end{eqnarray}
where $\mathcal{A}_{\dagger}$ and $\mathcal{A}$ are both column vectors, $\mathcal{A}_{\dagger}=(a^{\dagger}_1,a^{\dagger}_2,\cdots,a^{\dagger}_N)^t$ and $\mathcal{A}=(a_1,a_2,\cdots,a_N)^t$. $a_p$ ($a_p^{\dagger}$) is the annihilation (creation) operator of the $p$-th bare phonon mode.  $\mathbf{L}_g$ and $\mathbf{P}_g$ are diagonal matrices, whose matrix elements are $\mathbf{L}_g(p,p)=\sqrt{\hbar/2M\tilde{\omega}_p}$ and $\mathbf{P}_g(p,p)=i\sqrt{\hbar M\tilde{\omega}_p/2}$, where $\tilde{\omega}_p$ is the $p$-th phonon frequency of the bare mode. Similarly we obtains the result of normal coordinate $\mathcal{Q}_e$ and momentum $\mathcal{P}_e$ of the Rydberg shaped mode
\begin{eqnarray}
\label{seq:rydbergQ}
\mathcal{Q}_e&=&\mathbf{L}_e(\mathcal{B}_{\dagger}+\mathcal{B}),\\
\label{seq:rydbergP}
\mathcal{P}_e&=&\mathbf{P}_e(\mathcal{B}_{\dagger}-\mathcal{B}),
\end{eqnarray}
where $\mathcal{B}=(b_1,b_2,\cdots,b_N)^t$, $\mathbf{L}_e(p,q)=\sqrt{\hbar/2M{\omega}_p}$ and $\mathbf{P}_e(p,q)=i\sqrt{\hbar M\omega_p/2}$ when $p=q$ and zero otherwise.

We now find the Duschinsky transformation between the bare and Rydberg modes. The displacement of the ions around their equilibrium positions is denoted by a column vector $\mathcal{X}=(x_1,x_2\cdots,x_N)^t$ where $t$ indicates the transpose operation.  We can obtain that $\mathcal{Q}_g=\mathbf{A}\mathcal{X}$ and $\mathcal{Q}_e=\mathbf{B}\mathcal{X}$, where $\mathbf{A}$ is the eigenvector of the bare modes. Applying the Duschinsky transformation, one finds $\mathcal{Q}_e=\mathbf{T}\mathcal{Q}_g$ with $\mathbf{T}=\mathbf{AB}^{-1}$. Similarly one obtains $\mathcal{P}_e=\mathbf{T}\mathcal{P}_g$.

With Eqs.~(\ref{seq:bareQ})-(\ref{seq:rydbergP}) and the Duschinsky transformation, we can find the transformation between the phonon operators of the Rydberg and bare mode,
\begin{eqnarray}
\mathcal{B}_{\dagger}&=&\frac{1}{2}\left[\mathcal{T}_{+}\mathcal{A}_{\dagger}+\mathcal{T}_{-}\mathcal{A}\right],\nonumber\\
\mathcal{B}&=&\frac{1}{2}\left[\mathcal{T}_{-}\mathcal{A}_{\dagger}+\mathcal{T}_{-}\mathcal{A}\right],\nonumber
\label{eq:operator_transform}
\end{eqnarray}
where $\mathcal{T}_{\pm}=(\mathbf{L}_e^{-1}\mathbf{TL}_g\pm\mathbf{P}_e^{-1}\mathbf{TP}_g)$. This relation allows us to express the gate evolution operator Eq.~(\ref{eq:rydbergphonon_evolution}) in terms of the phonon operators of the bare mode,
\begin{eqnarray}
\label{eq:barephonon_evolution}
U(\tau)&=&\exp\left[-i(\mathbf{C}_g\mathcal{A}_{\dagger}+\mathbf{C}_g^*\mathcal{A})+i\sum_{mn}\phi_{mn}\sigma_m^z\sigma_n^z\right],\\
&=&\exp\left[-i\sum_p(\beta_pa^{\dagger}_p+\beta_p^*a_p)+i\sum_{mn}\phi_{mn}\sigma_m^z\sigma_n^z\right],\nonumber
\end{eqnarray}
where $\mathbf{C}_g=\text{Re}(\mathbf{C}_e)\mathbf{L}_e^{-1}\mathbf{TL}_g+i\text{Im}(\mathbf{C}_e)\mathbf{P}_e^{-1}\mathbf{TP}_g$  and $\beta_p=\sum_m\mathbf{C}_g(m,p)$. Eq.~(\ref{eq:barephonon_evolution}) permits us to calculate the gate fidelity conveniently as initially the bare phonon modes are in a thermal state. After some math, the gate fidelity is found,
\begin{eqnarray}
F&=&\frac{1}{256}\sum_{jk}\varphi_{jk}\nonumber\\
&&\times \exp\left[\frac{1}{2}\sum_p\left(\beta_p^j\beta_p^{k*}-\beta_p^{j*}\beta_p^k-|\beta_p^j-\beta_p^k|^2\coth\frac{\gamma_p}{2}\right)\right],\nonumber
\end{eqnarray}
with the temperature factor $\gamma_p=\hbar\omega_p/k_BT$ ($k_B$ the Boltzman constant, $T$ phonon temperature) and
\begin{eqnarray}
\varphi_{jk}&=&\exp\left[\frac{i\pi}{4}(\sigma_{m_1}^k\sigma_{n_1}^k-\sigma_{m_1}^j\sigma_{n_1}^{j}
+\sigma_{m_2}^k\sigma_{n_2}^k-\sigma_{m_2}^j\sigma_{n_2}^{j})\right.\nonumber\\
&+&\left.i\sum_{mn}(\phi_{mn}\sigma_m^j\sigma_n^j-\phi_{mn}^*\sigma_m^k\sigma_n^k)\right]. \nonumber
\end{eqnarray}
Here the eigenbasis of the 4-qubit state is $|j\rangle$, where $|j\rangle=|\uparrow,\uparrow,\uparrow,\uparrow\rangle,\cdots$ and $\sigma_{m}^{(z)}|j\rangle=\sigma_{m}^{j}|j\rangle$.

\end{document}